\documentclass[a4paper,11pt]{article}
\usepackage{jinstpub} 

\title{\boldmath Uncertainty Propagation within Chained Models for Machine Learning Reconstruction of Neutrino-LAr Interactions}

\author[a,1]{D. Douglas, \note{Corresponding author.}}
\author[a]{A. Mishra}
\author[b]{F. Petersen}
\author[a]{D. Ratner}
\author[a]{K. Terao}

\affiliation[a]{SLAC National Accelerator Laboratory,\\
Menlo Park, CA 94025}
\affiliation[b]{Stanford University,\\
Stanford, CA 94305}

\emailAdd{dougl215@slac.stanford.edu}

\abstract{Sequential or chained models are increasingly prevalent in machine learning for scientific applications, due to their flexibility and ease of development.
  Chained models are particularly useful when a task is separable into distinct steps with a hierarchy of meaningful intermediate representations.
  In reliability-critical tasks, it is important to quantify the confidence of model inferences.
  However, chained models pose an additional challenge for uncertainty quantification, especially when input uncertainties need to be propagated.
  In such cases, a fully uncertainty-aware chain of models is required, where each step accepts a probability distribution over the input space, and produces a probability distribution over the output space.
  In this work, we present a case study for adapting a single model within an existing chain, designed for reconstruction within neutrino-Argon interactions, developed for neutrino oscillation experiments such as MicroBooNE, ICARUS, and the future DUNE experiment.  
  We test the performance of an input uncertainty-enabled model against an uncertainty-blinded model using a method for generating synthetic noise.  By comparing these two, we assess the increase in inference quality achieved by exposing models to upstream uncertainty estimates.}


\begin{document}
\maketitle
\flushbottom

\section{\label{sec1} Introduction}

Deep learning has seen wide success and growing adoption in many fields of study and in various industries.
Increasingly, these techniques are employed in reliability-critical tasks, where a model's poor prediction can lead to grave outcomes, as in medical diagnostics~\cite{liu2022deep}, autonomous driving~\cite{grigorescu2020survey}, and other applications where human health may be implicated. 
Similarly, deep learning based modeling is also finding adoption in scientific applications, such as neutrino physics reconstruction~\cite{domine2020scalable}, control of particle accelerators~\cite{gupta2021improving}, etc. 
In such fields, erroneous models can lead to deleterious effects on scientific knowledge and progress.

For such applications, in addition to the model's predictions, we require a measure of the model's confidence in its predictions, e.g., in the form of prediction intervals, calibrated probabilities, etc. 
One method for increasing the reliability of predictions is to produce a measure of confidence associated with a given model prediction.
This additional information, along with the most probable or mean value, can enable a user to selectively ignore anomalous inferences, synthesize confident and uncertain inferences in a holistic way, and help diagnose mis-modeling for certain inputs. 

Uncertainty in a model's prediction can arise due to several sources~\cite{kendall2017uncertainties, tagasovska2019single}, typically categorized into \textit{epistemic} and \textit{aleatoric}.
The choice of model's architecture and complexity defines the space of possible functions which can be expressed, and the disparity between this space and the true process being modeled represents the model's epistemic uncertainty~\cite{antoran2020depth}. 
Once the structure of a model is decided, it is then trained using a sample drawn from a distribution of possible inputs for the task.  
In this training process, a second type of uncertainty -- aleatoric uncertainty -- is introduced, through the disagreement between the training set and the real input space, as well as approximation error accumulated by fitting model weights numerically.  
In some cases, the input itself may have uncertainty, as it may be the product of a physical measurement, or the output of another uncertainty-enabled model. In this case, a model should propagate the input as a distribution, predicting an output distribution whose shape is a derived from both the input and model uncertainties.

With the advent of end-to-end deep learning approaches, most applications use a single model accepting data, carrying out representation learning, and outputting the final predictions~\cite{amodei2016deep}. 
However, in certain cases, it is desirable to approach the problem as a set of multiple discrete tasks in a sequence. 
This can be particularly useful when there are meaningful representations of the data between the input and final output spaces. 
These intermediate representations may be physically interesting in and of themselves.  They may also correspond to an independently measurable quantity that can be used to form a specialized training or validation set for a separable task in the chain. 
In such a chain of models, errors and uncertainties from an upstream model in the sequence can affect the performance of the downstream models. 
Testing the performance of any of the individual models in the chain in isolation may not be an adequate measure of their final performance when deployed in the chain.
Thus, it is essential to ascertain the effects of the propagation of uncertainties from the upstream models to the downstream models at each stage in the chain.
Additionally, it is essential to understand the dependence of the final uncertainty upon the uncertainty of inputs, for various kinds of inputs with varying degrees of measurement error, including which steps in the chain have an inflationary or reductionary effect.

Chains of data driven models have been applied to complex tasks in material science~\cite{shafighfard2024chained}, fluid mechanics~\cite{wahid2023multiphase}, etc. Furthermore, they are often used in autonomous driving applications. Le et al.~\cite{le2023comparing} compared the performance of chained neural networks to an end-to-end deep learning model for the task of locating nearest objects in geographic images. 
They found that while both approaches had similar accuracy, chained composite models exhibited an order of magnitude reduction in computational expense for training.

In this paper, we will focus on a sequential model which was developed for the reconstruction of neutrino-argon scattering events in a Liquid Argon Time Projection Chamber (LArTPC), a detector design widely used in neutrino oscillation experiments, such as the future Deep Underground Neutrino Experiment (DUNE).
The initial input for this model, the detector output, is represented by a sparse point cloud -- measurements of ionization signals left in the wake of energetic charged particles interacting in the active liquid argon (LAr) medium.  
The reconstruction model first performs image-based tasks, including semantic classification of voxels and location of key points like potential vertices, start and end points of track-like fragments.  
Further downstream models represent detector outputs as an abstracted particle flow description, reconstructing the physical type and kinematic properties of each particle as energy is deposited in the sensitive volume over the course of the interaction's timeline.

A more complete description of this model can be found in Section~\ref{sec:SPINE}.  In the context of this reconstruction chain, we study the effects of uncertainty propagation in chained machine learning models.
We compare the accuracy and uncertainty predictions of both uncertainty-aware (UA) and uncertainty-blinded models (referred to as ``blind''), by producing parallel models trained with and without per-feature uncertainty estimates.  
We demonstrate improvements in these metrics in the context of the node and edge classification tasks performed by a Graph Neural Network (GNN) (the first model labelled as ''GrapPA`` in Figure \ref{fig:mlreco}).  

\section{\label{sec2} Background: Liquid Argon Time Projection Chambers and ML Reconstruction}


\subsection{The Liquid Argon Time Projection Chamber}

\begin{figure}
  \centering
  \includegraphics[width=0.75\textwidth]{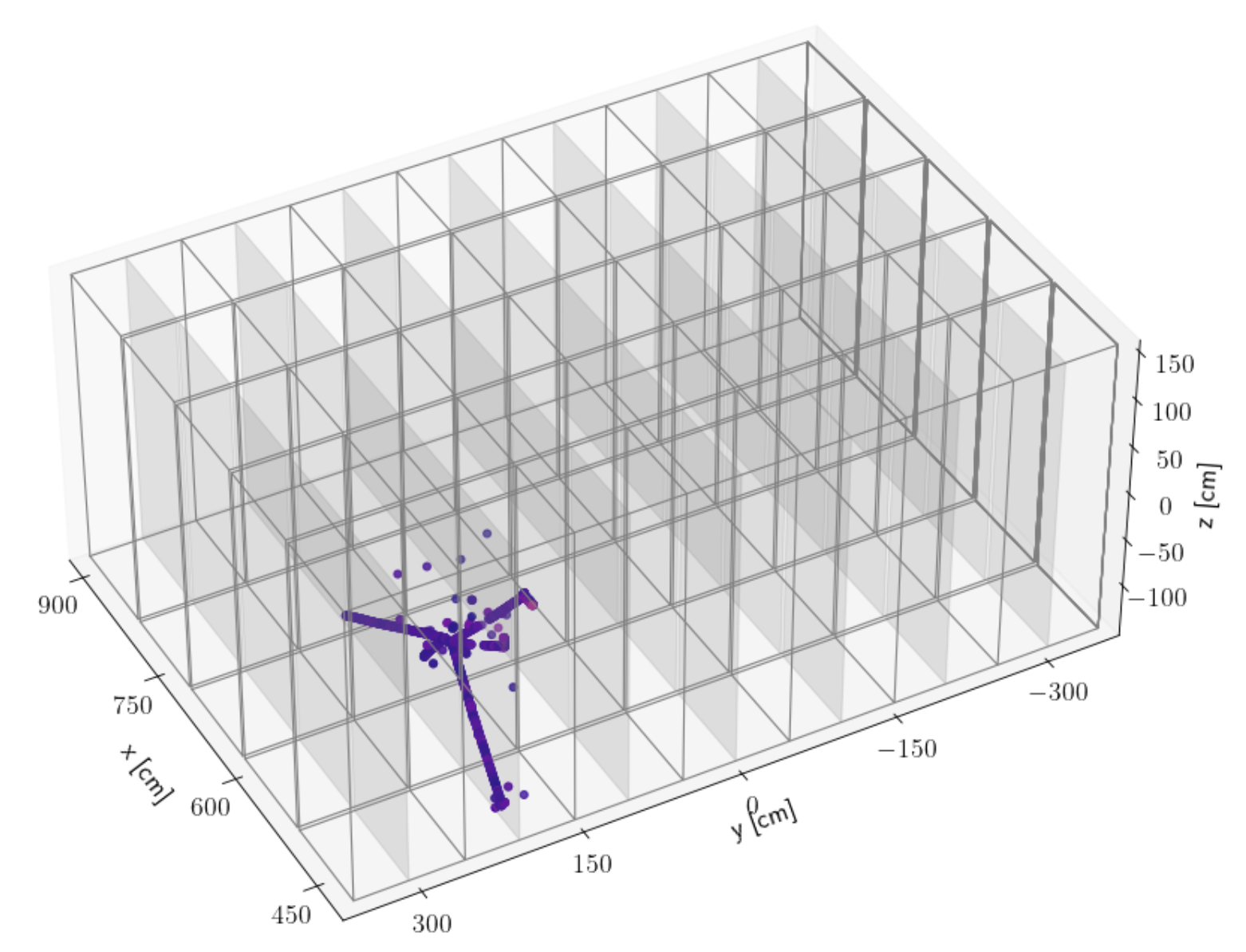}
  \caption{\label{fig:example_event_display} An example of a simulated neutrino interaction in the DUNE ND geometry, displaying typical degree of sparsity.  This detector features a highly-segmented design, consisting of 35 dual-TPC modules packed in a 5 $\times$ 7 grid.}
\end{figure}

A LArTPC is a 2D/3D imaging detector technology used prominently in neutrino physics experiments because of their high quality spatial and energy resolution. 
A large volume of LAr acts as both the target for incoming neutrinos and the detection medium for imaging. 
When a neutrino interacts on an Argon nucleus within the detector, energetic charged particles are created at the interaction point, or \textit{vertex}. These secondary particles continue through the material, leaving trails of ionization as they propagate through the material.  
Figure~\ref{fig:example_event_display} shows an example of a neutrino interaction, with points representing the location of ionization signals within the detector. 

In this study, we simulated the current design of the Deep Underground Neutrino Experiment (DUNE) near detector. Here, the imaging volume is 5~m $\times$~7~m~$\times$~3~m recorded in the 3D image of a dimension 1\,250~$\times$~1\,750~$\times$~1\,000 voxels.

A typical neutrino interaction may produce several particles with varying size including a straight proton trajectory of a few centimeters, a meters-long muon and/or charged pions as well as electromagnetic particles such as electrons each of which produces a {\it fragmented} trajectory (i.e.\ multiple disconnected short trajectories).  Each voxel represents a region in which charge ionized by these particles may be measured.  When charge meets a pre-configured threshold, it is digitized and saved.  Thus, a typical interaction captured by the LArTPC design is densely sampled in the signal region (i.e.\ particle trajectories) without gaps for high precision physics inference, but is otherwise globally sparse.  

\subsection{Machine Learning-based Reconstruction of LArTPC Events (SPINE Reconstruction Model)}

\label{sec:SPINE}

\begin{figure*}
  \centering
  \includegraphics[width = \textwidth]{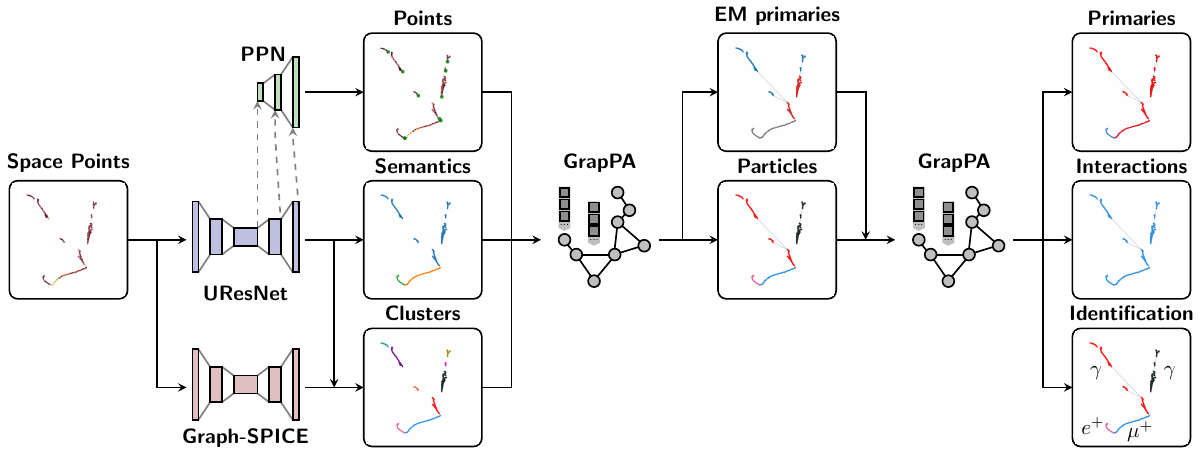}
  \caption{\label{fig:mlreco} Machine learning-based reconstruction chain for neutrino interaction analysis in a LArTPC \cite{drielsma2021scalableendtoenddeeplearningbaseddata}.  In this paper, we examine the first model labelled as ``GrapPA'', in which disconnected shower fragments are partitioned and initial fragments are identified.}
\end{figure*}

Scalable Particle Imaging with Neural Embedding (SPINE)~\cite{drielsma2021scalableendtoenddeeplearningbaseddata} is the current state-of-the-art technique for analyzing LArTPC detector data. At its core, SPINE is a chain of multiple convolutional (CNNs) and graph neural networks which infer hierarchical physics features from the low to high level in sequential manner.  The chain shown in Figure~\ref{fig:mlreco} can be broken into two parts: the first stage utilizes CNN-based algorithms to infer pixel-level information. The output of the first stage is {\it particle fragments} which are essentially clusters of pixels. Each fragment contains pixels that are densely connected (i.e.\ no gap) and several fragment-level information are also inferred using CNNs. For physics analysis, we are interested in individual particle instance which consists of one or more of these fragments. 

The second stage of SPINE utilizes GNNs, noted as GrapPA in Figure~\ref{fig:mlreco}, with the goal of identifying the right set of fragments that represent individual particle instance. Our study targets the GrapPA GNN model. The input graph is constructed by interpreting indvidual particle fragment from the output of the first stage as a node. The edges are constructed between all possible pairs, thus the input is a fully-connected graph. Then the edge/node regression/classification is applied to infer features. For instance, the edge classification is used to infer the right set of nodes (fragments) to infer a particle instance, and the node classification for classifying different types of particle fragments.

The input to the GrapPA stage is the fixed number of node and edge features which introduce a bottleneck of information propagation as they must represent features associated with an arbitrary number of pixels in the corresponding cluster. Therefore the uncertainties of input become critical to inform how accurately key features are captured. Furthermore, unlike the pixel-level features extracted by prior CNNs, GrapPA is the first stage in the chain to produce the outputs that are used by a wide range of physics analyses that follow after SPINE. Predicting the output uncertainty at the GrapPA stage will hence provide more informed outputs and can make an impact in physics measurements.

\section{Computational Approach}

\subsection{Literature Review}
Numerous papers have studied the application of deep learning models for LArTPCs, including the reconstruction chain discussed above. Several of these have included studies of uncertainties for single-stage deep learning reconstruction-chain models; example tasks were single-particle classification, multi-particle classification, and semantic segmentation. \cite{buuck2023low, khek2022gamma} compared different deep learning UQ approaches for epistemic uncertainties in the initial positions and directions of electrons scattered in Compton interactions in a low-energy LArTPC gamma ray telescope. \cite{Abratenko_2022} investigated the influence of uncertainties and errors in the detector response on the MicroBooNE physics analyses. This was carried out using an approach which is agnostic toward any specific Machine Learning model used for the reconstruction task. \cite{koh2023deep, koh2021evaluating} evaluated deep learning UQ methods on the task of particle classification using the PiLArNet Monte Carlo 3D LArTPC point cloud dataset \cite{adams2020pilarnet}. However, while prior investigations have analyzed the epistemic and aleatoric uncertainties for single stages in the LArTPC reconstruction chain, there is a need to understand the propagation of input uncertainties through different stages in the reconstruction chain.

Outside of neutrino physics, various papers have studied the propagation of uncertainties through neural networks. Examples include variants of Moment Matching techniques \cite{wang2013fast} and Stable Distribution Propagation \cite{petersen2024uncertainty}. \cite{postels2019sampling} showed a sampling-independent approach to propagating epistemic uncertainties in in autonomous driving based semantic segmentation on the CamVid dataset. For chained models, \cite{shao2024uncertainty} formulated motion planning for autonomous driving as a continuous space, discrete-time model, where the first model stage predicts the future motion of other traffic participants and the ensuing model stage uses this input to determine the vehicle's trajectory. Their results underscored the advantages of integrating uncertainties from prior stages in the motion planning task. Similarly, in autonomous vehicle motion planning, \cite{ivanovicimportance} have analyzed the need for calibrated uncertainties from upstream models for motion planning. In their analysis, the authors observed that better calibrated upstream predictions, led to improved metrics for the uncertainty-aware planner stage.

\subsection{Method}
\label{sec:Method}
In this study, we investigate how an uncertainty-enabled model within a large chain of such models makes well-calibrated inferences.  
We focus on Probabilistic Neural Networks (PNNs), which in this case means a model which predicts the parameters of a distribution over the prediction space.  
For a regression problem with Gaussian errors, this distribution may follow a normal distribution, while a 2-class classification problem may be expressed with a Bernoulli probability.  
A model should also be able to ingest the distributional description of an input and learn the connection between the shape of the input distribution and the confidence of the input.

\begin{figure}
  \centering
  \includegraphics[width=\textwidth]{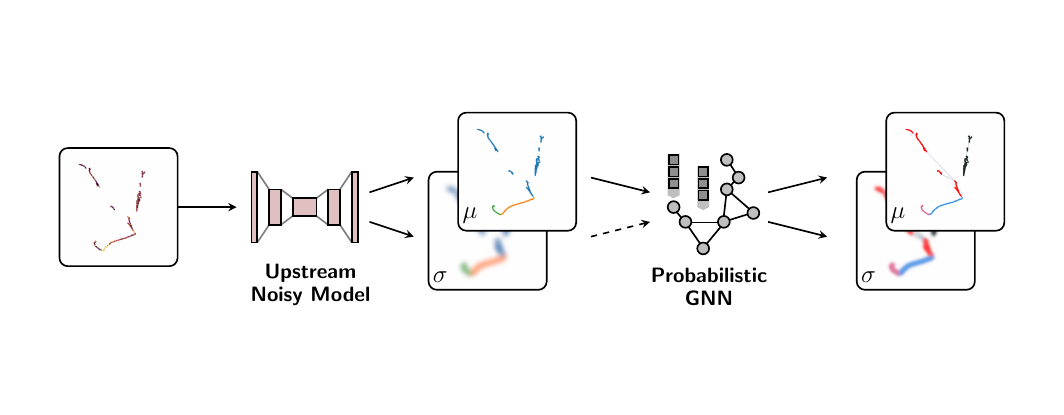} 
  \caption{\label{fig:blinding_schematic} Testing scheme for the performance of a chained model with and without access to upstream uncertainty information.  A model under consideration (Here, ``Probabilistic GNN'') recieves a noisy inference from an upstream model in the sequence.  In the uncertainty-aware scheme, this model also sees a well-calibrated estimate of the uncertainty associated with the noise of the upstream model (dashed arrow).  In the case of a blinded model, the upstream uncertainty estimate is replaced with a static value, masking the information, but retaining the shape of the input, so as to leave the model's structure unaltered.  Our hypothesis is that the uncertainty-aware model will always out-perform the blinded model either in terms of accuracy or in terms of improved confidence estimates.}
\end{figure}

To assess the quality of the uncertainty estimates, we construct a blinding test (Figure \ref{fig:blinding_schematic}): we train two models identical in structure on the same input set, with one version receiving the full distributional information (mean and width parameters) as input features, while the other model receives only the mean information features.  This second, ``blind'' model represents what a model with noisy deterministic inputs may hope to achieve, while the ``uncertainty-aware'' model can in principle make better predictions in terms of both accuracy and calibration of output uncertainties, without changing model complexity.  As the two models share the same architecture, the blind model is given the same number of input features, but the input uncertainty features are masked with a static value.

\begin{align*}
    f_\mathrm{Blind} &: \hat{x} \rightarrow y, \sigma_y \\
    f_\mathrm{Uncertainty-Aware} &: \hat{x}, \sigma_{\hat{x}} \rightarrow y, \sigma_y
\end{align*}

In our case study, we do not have access to true upstream uncertainties, but can instead construct some noisy inputs by adding synthetic noise.  We first choose a scale of uncertainty -- 10\% - 30\% of the ground truth value in our example -- based on our expected uncertainty output from the upstream step.  Then, synthetic noise is added to the deterministic features as follows:

\begin{align*}
  \sigma_i &= \alpha_i x_i \\
  \alpha_i &\sim U[0.1, 0.3] \\
  \hat{x}_i &= x_i + \beta_i \\
  \beta &\sim N[0, \sigma_i]
\end{align*}

Using this method for per-feature smearing, we obtain an $\hat{f}_i$ and $\sigma_i$ per deterministic feature where the noise is normally distributed and $\sigma_i$ gives an estimate of the error distribution, but not the true error $|\hat{f}_i - f_i|$.

The two approaches are trained on simulated data using a binary cross-entropy loss function.

In discussions of model performance, we will focus on the metrics of accuracy, binary cross-entropy, mis-calibration area, and binary score entropy.  Accuracy here is the observed frequency of correctly-labelled classification items.  Mis-calibration area (MCA) is an assessment of the correlation between the true error rate and the expected error rate given the model's predicted uncertainty.  Deviation of this metric from unity represents an inability of the predicted uncertainties to adequately map to the true error, and so the degree of this disagreement is represented by the area between the model's uncertainty and the ideal predictor \cite{Calibration}.  Binary entropy here is the entropy of the model's output treated as a Bernoulli process \cite{mackay2003information}.  This metric gives an estimate of the sharpness of the model's predictions, as it indicates the likelihood of scores to be preferentially in the high-confidence regions -- near 0 or 1, producing a low entropy value -- or in the low-confidence regions -- near the decision boundary, producing a high entropy value.

\section{\label{sec5} Results \& Discussion}

As an example for adding uncertainty quantification to the existing machine learning-based reconstruction chain, we examine the first GNN stage outlined in Figure \ref{fig:mlreco}, labelled ``GrapPA''.  At this stage in the reconstruction of a LArTPC image, it is common to find many independent interactions which are coincident in time. Each of the showers displays a high level of fragmentation, due to the prevalence of neutral particles which may be produced in one region of the detector and travel for some distance before re-interacting to again produce visible, charged particles \cite{PhysRevD.104.072004}.

The tasks at this stage are two-fold: to associate each fragment to the other fragments belonging to the same shower, and to identify the most upstream ``parent'' fragment within each shower. For this problem, the inputs are presented as a fully-connected graph, where each fragment is represented as a node with a feature vector whose values are derived from the fragment's point cloud representation:

\begin{itemize}
    \item{Fragment point cloud centroid (3 parameters)}
    \item{Fragment point cloud covariance (upper diagonal) (6 parameters)}
    \item{Fragment size (number of points) (1 parameter)}
\end{itemize}

Likewise, each edge in the fully-connected graph is represented by a feature vector which is defined by the two point cloud representations of the fragments it connects:

\begin{itemize}
    \item{Points of closest approach (6 parameters)}
\end{itemize}

As described in Section \ref{sec:Method}, blinded and uncertainty-aware models are trained on simulated data of electromagnetic showers using a generic event generator \cite{PhysRevD.104.072004}.  Independent shower events are simulated and superimposed within the detector geometry to produce a data-like sample with a high degree of pileup.

This model is designed to infer both the node states and the edge states of the graph, and simultaneously predict a label for each of these tasks.  In the \textit{edge task}, the label indicates whether an edge connects two causally-related showers, so that all pairs of fragments within a shower will share an edge with label 1.  Conversely, edges conneting pairs of fragments belonging to different showers are labelled 0.  Second, the \textit{node task} is a per-node label inference, where a true label of 1 corresponds to the fragment which initiated the larger shower structure, while downstream, ``child'' nodes have a true label of 0.

There are some significant qualitative differences in these tasks.  The classes represented in the node task are inherently unbalanced, with one parent node corresponding to many child nodes (give an actual estimate of this imbalance).  Likewise, the prevalence of truly ``on'' edges is much smaller than that of truly ``off'' edges in the edge classification task.  Lastly, properly associating all of the fragments within a shower without including is more complicated than just classifying edges as ``on'' or ``off'', as partitioning of the graph into its proper sub-graphs involves ensuring that all edges connecting unrelated showers are predicted as ``off''.  In the full ML-reco scheme, this is done with an adjustable cut, using an Adjusted Rand Index (ARI) metric in constructing each sub-graph \cite{PhysRevD.104.072004}.

The node classification task can be seen as somewhat simpler, in that the shape of each parent node will often present a ``cone'' which will point towards child nodes.  It is observed in both the blinded and uncertainty-aware models that the node task is able to achieve a higher baseline accuracy than the edge task for each trained model.  This logic can be applied to the edge classification task, to the extent that this effect will make classification of edges between parent nodes and their child nodes simpler, but adds no benefit for predicting the true label between child nodes within a shower and across unrelated showers.

\begin{figure}
  \centering
  \fbox{\includegraphics[width=0.45\linewidth, trim={3.5cm 1.5cm 4cm 1.7cm}, clip]{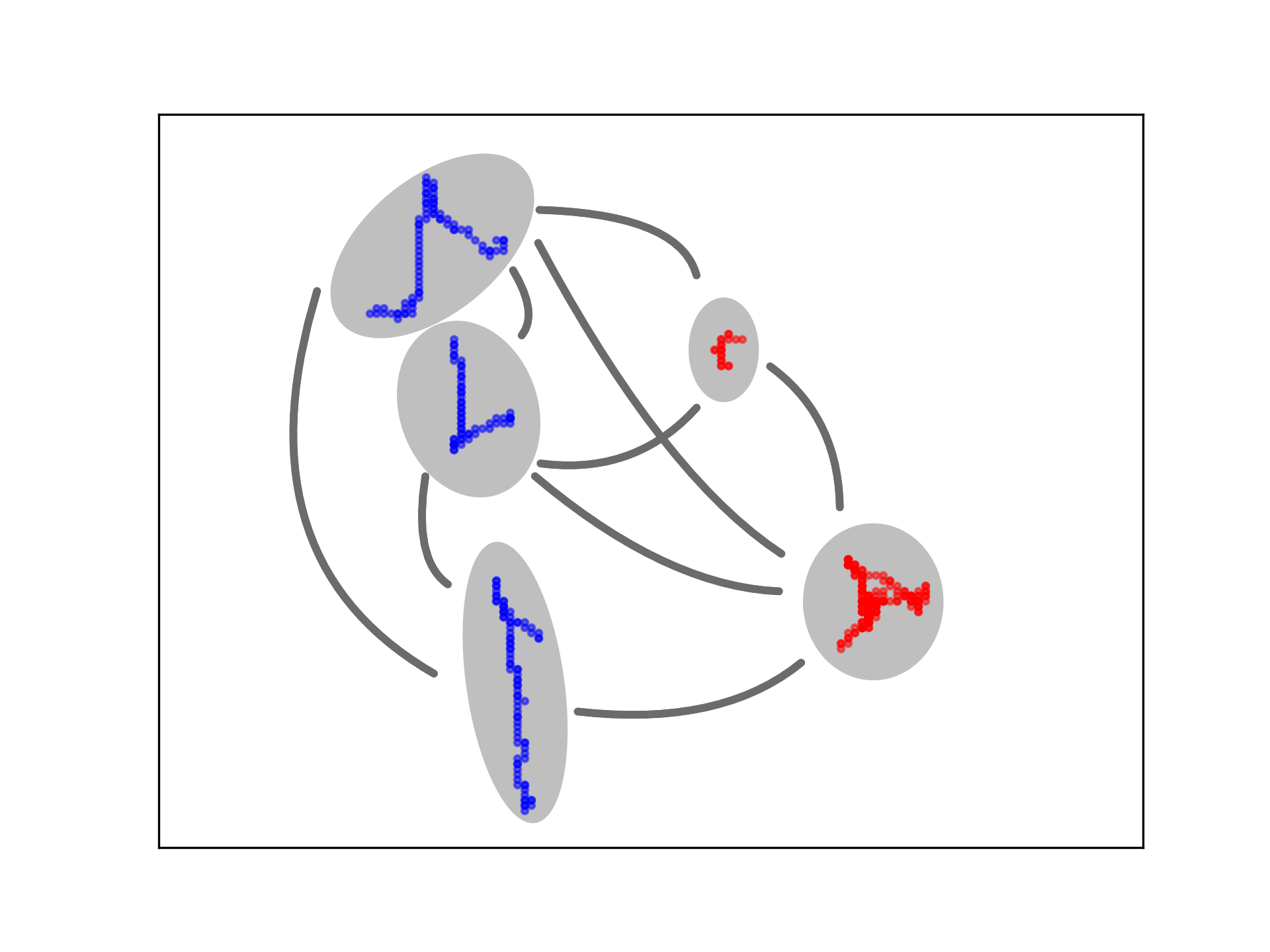}}
  \caption{\label{fig:graphexample} An example input for the shower fragment association task.  Here, 2 independent shower objects are represented by 5 fragments.  The point-cloud representation of each fragment is used to form a feature vector of uniform length, and each pair of fragment point clouds are likewise aggregated into a vector of fixed length.  In this way, the fragmented point cloud images are represented as a fully-connected graph.}
\end{figure}

For a set of models trained with an injected noise between 10\% and 30\%, the resulting metrics are summarized in Table \ref{tab:metrics}.  

\begin{figure}
  \centering
  \includegraphics[width=0.75\linewidth]{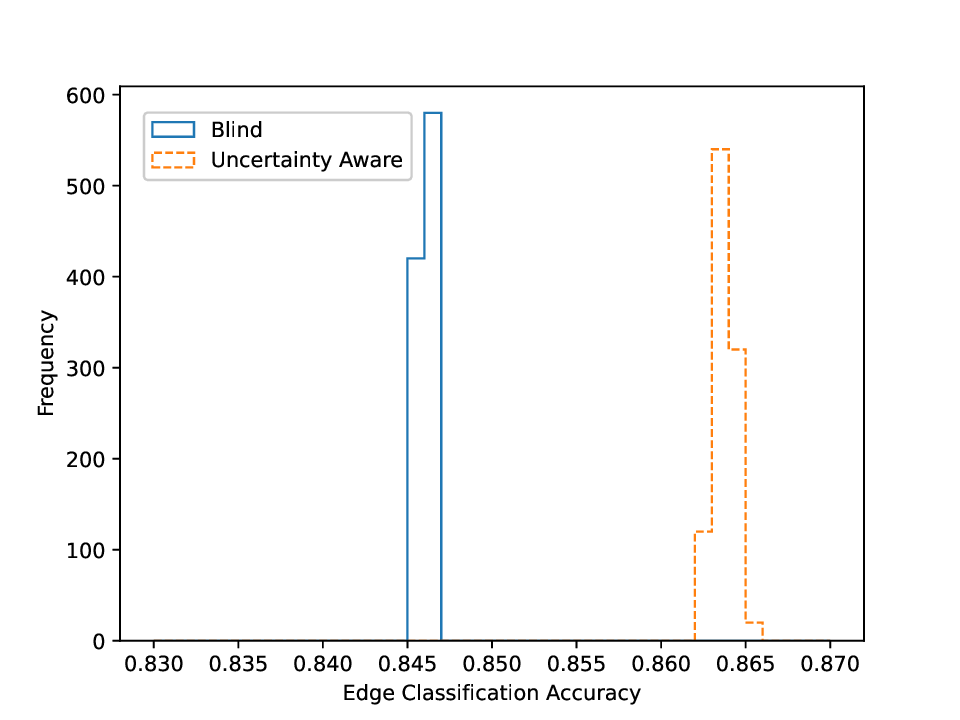}
  \caption{\label{fig:gnnEdgeAcc} Edge classification performance of the blinded GNN model compared to the uncertainty-aware model.  For this task, we see a statistically significant improvement by exposing the model to input uncertainty information.}
\end{figure}

\begin{figure}
  \centering
  \includegraphics[width=0.75\linewidth]{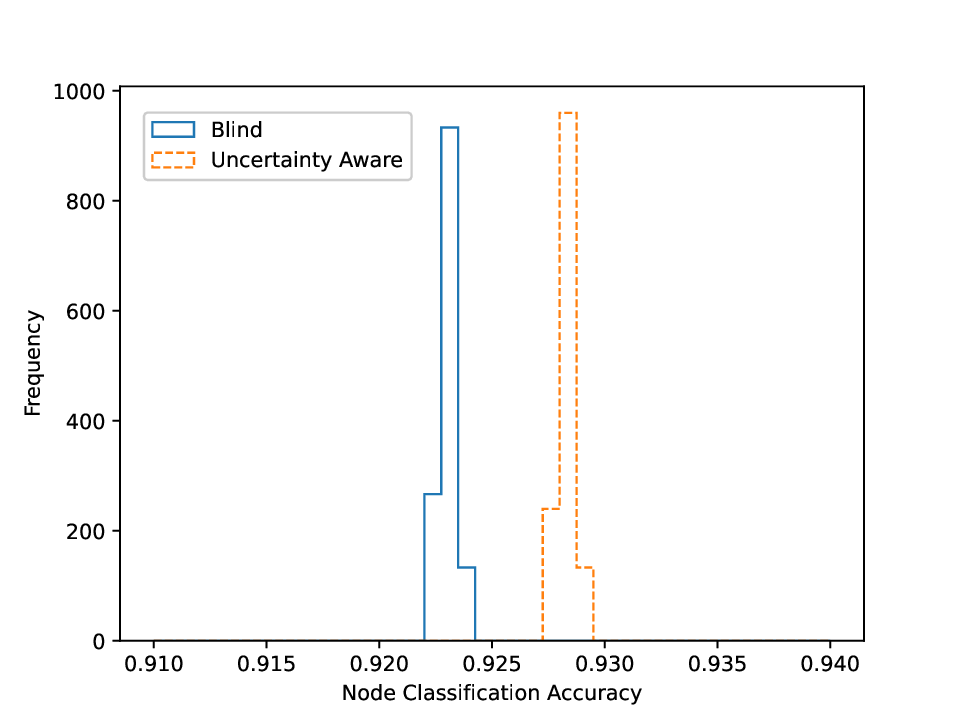}
  \caption{\label{fig:gnnNodeAcc} Node classification performance of the blinded GNN model compared to the uncertainty-aware model.  For this task, we see a smaller improvement when adding uncertainty information than is observed in the edge task.}
\end{figure}

We see a marked separation in the accuracy of blinded vs. unblinded models for the edge classification task, shown in Figure \ref{fig:gnnEdgeAcc}.  In the node classification task, shown in Figure \ref{fig:gnnNodeAcc} there is much less separation in performance, though there is a better overall accuracy when compared to the edge task.

\begin{table*}
\begin{tabular}{c|c|c|c|c}
    Metric & Accuracy & Loss & MCA & Entropy \\
    \hline
    Blind (edge task) & \; 0.846 $\pm$ 0.001 \; & \; 0.333 $\pm$ 0.001 \; & \; 0.0135 $\pm$ 0.0074 \; & \; 0.482 $\pm$ 0.006 \; \\
    UA (edge task) & 0.864 $\pm$ 0.001 & 0.301 $\pm$ 0.001 & 0.0161 $\pm$ 0.0091 & 0.437 $\pm$ 0.006 \\
    Blind (node task) & 0.923 $\pm$ 0.000 & 0.188 $\pm$ 0.000 & 0.0109 $\pm$ 0.0029 & 0.275 $\pm$ 0.005 \\
    UA (node task) & 0.928 $\pm$ 0.000 & 0.176 $\pm$ 0.000 & 0.0117 $\pm$ 0.0038 & 0.256 $\pm$ 0.005 \\
\end{tabular}
\caption{\label{tab:metrics} Metrics evaluated on an ensemble of 50 models each of the Uncertainty-Aware (UA) and Uncertainty-Blinded (blind) models.  See Figures \ref{fig:gnnEdgeAcc} through \ref{fig:gnnNodeEntropy}}
\end{table*}

\begin{figure}
  \centering
  \includegraphics[width=0.75\linewidth]{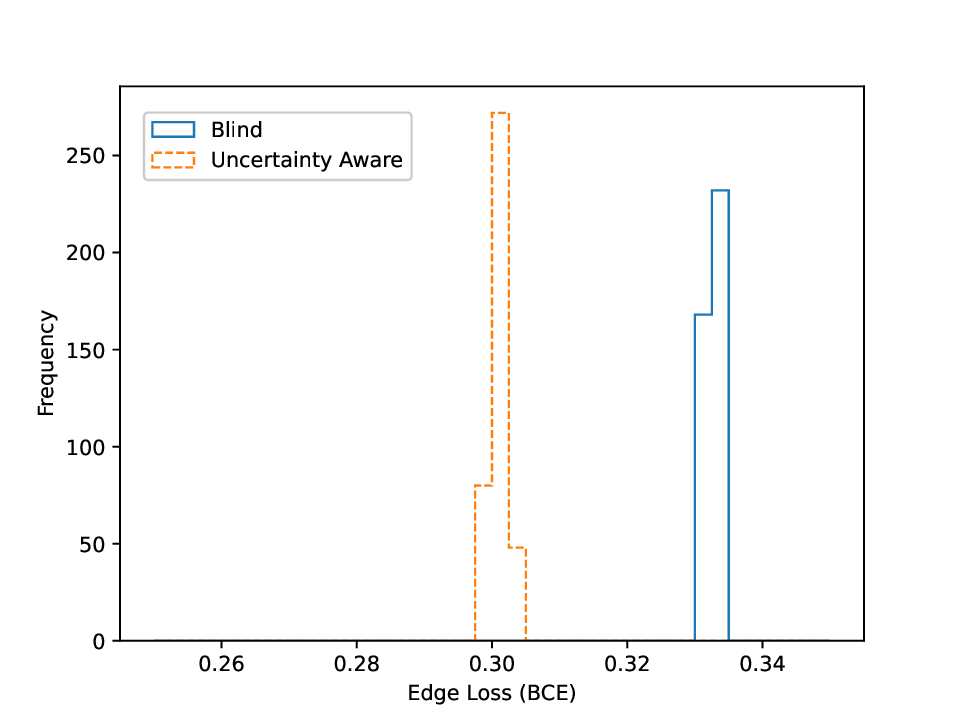}
  \caption{\label{fig:gnnEdgeLoss} Edge classification loss distribution for the ensemble. The loss here is binary cross-entropy, which corresponds to the negative log-likelihood for a Bernoulli process.}
\end{figure}

\begin{figure}
  \centering
  \includegraphics[width=0.75\linewidth]{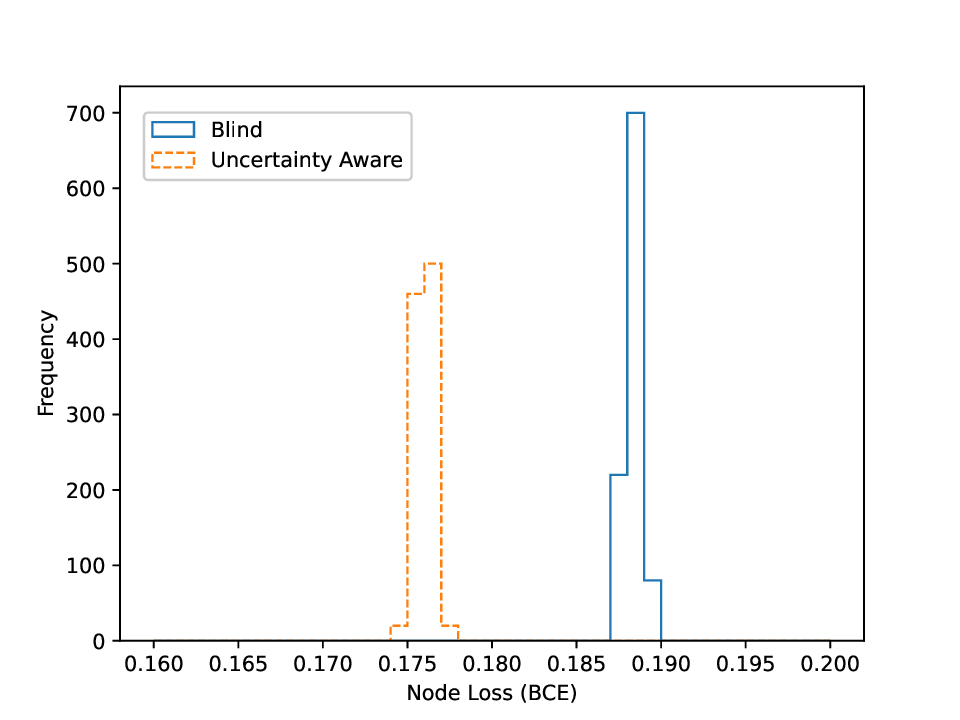}
  \caption{\label{fig:gnnNodeLoss} Node classification loss distribution for the ensemble.}
\end{figure}

There is also a significant difference in the final loss achieved by the two kinds of models.  We see an improvement in loss for both the edge task and the node task in Figures \ref{fig:gnnEdgeLoss} and \ref{fig:gnnNodeLoss}.

\begin{figure}
  \centering
  \includegraphics[width=0.75\linewidth]{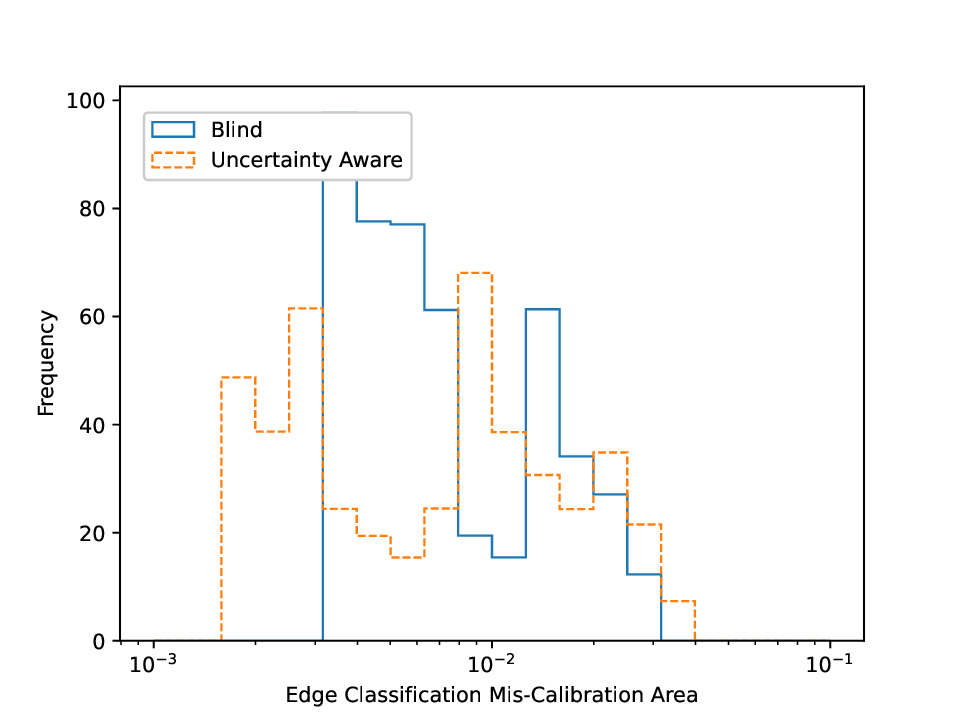}
  \caption{\label{fig:gnnEdgeMCA} Edge mis-calibration area for the ensemble.}
\end{figure}

\begin{figure}
  \centering
  \includegraphics[width=0.75\linewidth]{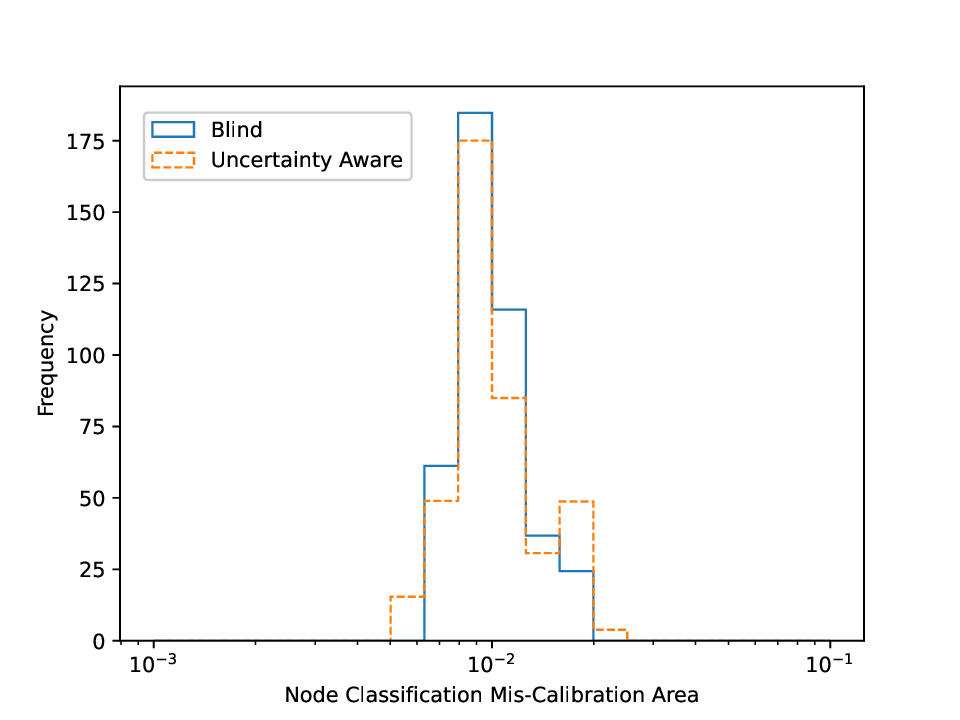}
  \caption{\label{fig:gnnNodeMCA} Node mis-calibration area for the ensemble. Both kinds of models produce nearly identically well-calibrated inferences.}
\end{figure}

Interestingly, both models produce equally well-calibrated scores with no additional methods applied to enforce this.  Figure \ref{fig:gnnEdgeMCA} shows broadly similar distributions for both kinds of models, with some individual uncertainty-aware models significantly out-performing the blind models.  Figure \ref{fig:gnnNodeMCA} shows a nearly identical result for the two kinds of model when performing the node task.  


Lastly, as a measure of sharpness, or a model's ability to make more confident predictions, we use the binary entropy of each model's predictions.  This metric yields a lower value for score distributions which are peaked near the true label values -- corresponding to high-confidence inferences -- while it yields high values when inferences are clustered near the decision boundary -- indicating low-confidence.  This metric is constrained to the [0, 1] interval.  What is observed is a significantly higher clustering of scores in the regions of high-confidence for the uncertainty-aware models in both tasks, with the edge task exhibiting a more pronounced separation.

\begin{figure}
  \centering
  \includegraphics[width=0.75\linewidth]{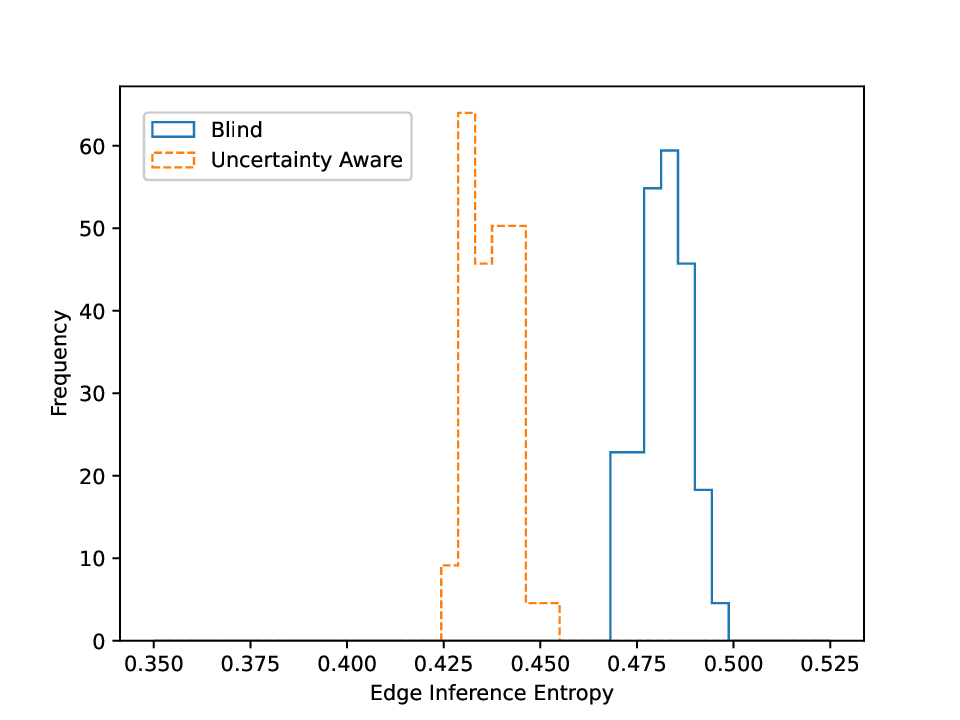}
  \caption{\label{fig:gnnNodeEntropy} Score binary entropy distribution for the edge task.  Uncertainty-Aware models show a systematically lower value, indicating higher clustering towards the true label values, and more confident predictions.}
\end{figure}


\begin{figure}
  \centering
  \includegraphics[width=0.75\linewidth]{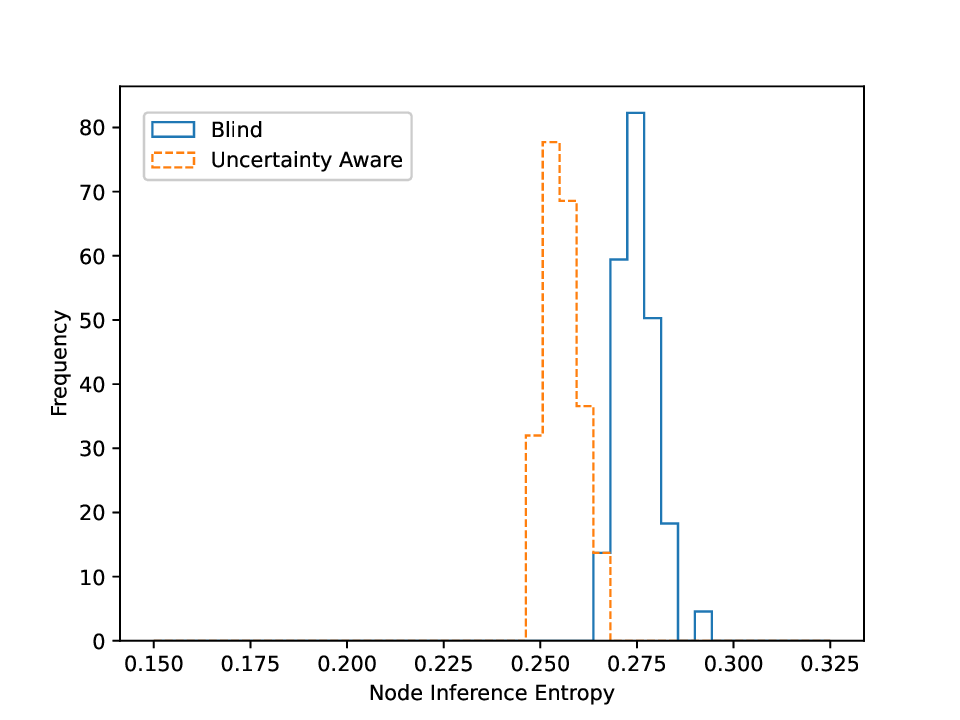}
  \caption{\label{fig:gnnEdgeEntropy} Score binary entropy distribution for the edge task.  Uncertainty-Aware models show a systematically lower value, indicating higher clustering towards the true label values, and more confident predictions.}
\end{figure}

\begin{figure}
  \centering
  \includegraphics[width=\linewidth]{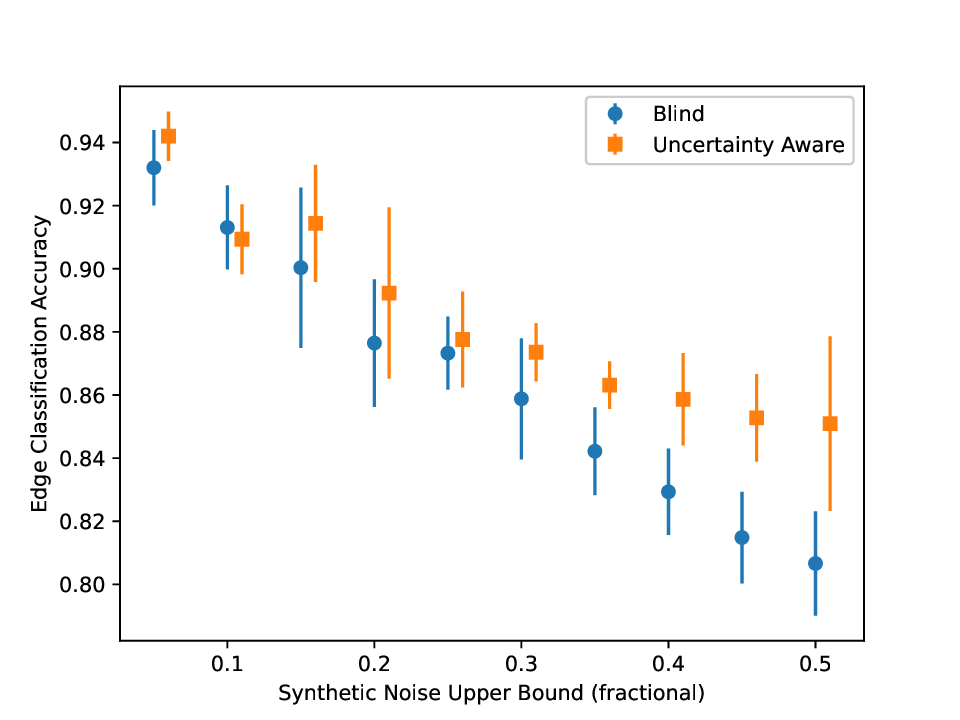}
  \caption{Improvement in accuracy of edge classification from a blinded model to an uncertainty-aware model for a range of injected feature noise.
\label{fig:gnnEdgeScan}}
\end{figure}

\begin{figure}
  \centering
  \includegraphics[width=\linewidth]{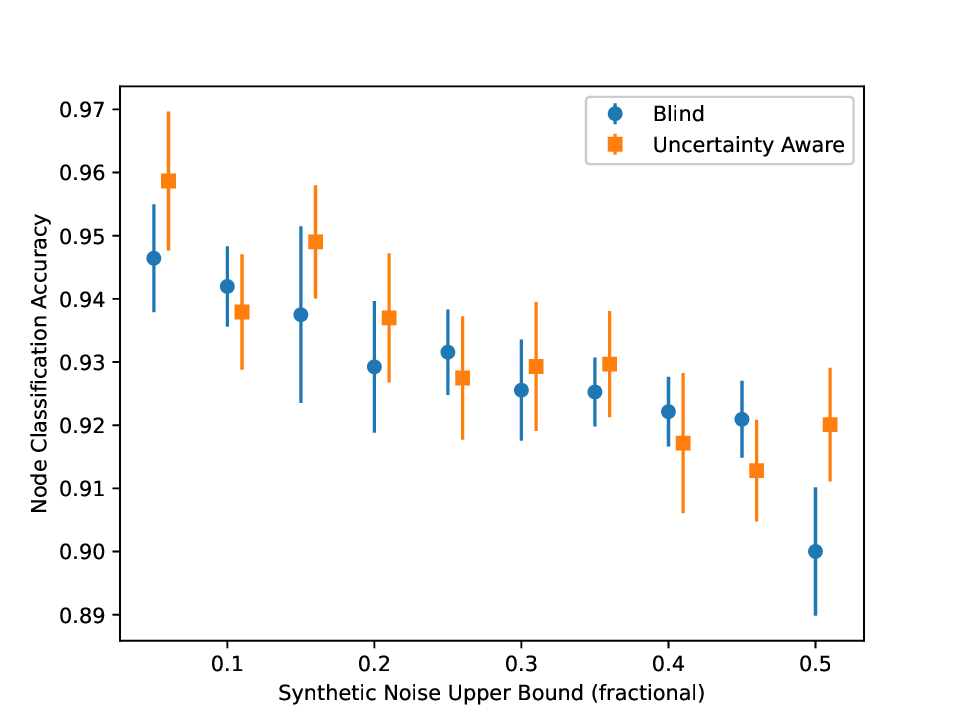}
  \caption{Node classification performance of the blinded GNN model compared to the uncertainty-aware model for a range of injected feature noise.  Note that this task does not see the improvements observed in the edge classification task.
\label{fig:gnnNodeScan}}
\end{figure}

As the gain in improvement is expected to depend strongly upon the level of noise in the input features, we also examined models trained on noise at different scales.  Figures \ref{fig:gnnEdgeScan} and \ref{fig:gnnNodeScan} show the comparative accuracy of blind and uncertainty-aware models with input uncertainties ranging from 5\% (fixed) to 5\% - 50\% per-feature.  From this scan, we start to see a separation in performance between the blind and uncertainty-aware models around 5\%-15\%, while the node classification task shows a comparable level of performance throughout the range we probed.

\section{\label{sec6} Conclusions}

The objective of this study is to investigate if in a sequence of chained models, having access to predictive uncertainties from upstream models led to improvements in the performance of the downstream models. This improvement could be quantified via improved accuracy, better calibration, or increased confidence of predictions.

We have seen that for models embedded within a larger chain, as are an increasingly common approach in complex tasks, there is a need for uncertainty quantification in terms of model noise, and in terms of propagation of per-input uncertainties to output uncertainty.

In this paper, we have examined a GNN-based problem within a neutrino-argon scattering reconstruction algorithm, and observed that edge inference is greatly improved for noisy labels when a model is trained with an estimate of the input uncertainty.  In this same case study, we have also seen that the node classification task is less sensitive to noise added in this way, as both the blinded and uncertainty-aware models perform approximately equally well for a wide range of noise parameters.

\begin{acknowledgments}
This work was supported by the U.S. Department of Energy, Office of Science, Office of High Energy Physics under Contract DE-AC02-76SF00515. Computational resources were provided by the SLAC Shared Scientific Data Facility.
This work was also supported in part by Zoox, Inc.
\end{acknowledgments}

\bibliographystyle{JHEP}
\bibliography{uq_neutrino.bib}

\providecommand{\noopsort}[1]{}\providecommand{\singleletter}[1]{#1}%

\providecommand{\href}[2]{#2}\begingroup\raggedright\begin{thebibliography}{10}

\bibitem{liu2022deep}
T.~Liu, E.~Siegel and D.~Shen, \emph{Deep learning and medical image analysis
  for covid-19 diagnosis and prediction}, {\emph{Annual review of biomedical
  engineering} {\bfseries 24} (2022) 179}.

\bibitem{grigorescu2020survey}
S.~Grigorescu, B.~Trasnea, T.~Cocias and G.~Macesanu, \emph{A survey of deep
  learning techniques for autonomous driving}, {\emph{Journal of field
  robotics} {\bfseries 37} (2020) 362}.

\bibitem{domine2020scalable}
L.~Domin{\'e}, K.~Terao and D.~Collaboration), \emph{Scalable deep
  convolutional neural networks for sparse, locally dense liquid argon time
  projection chamber data}, {\emph{Physical Review D} {\bfseries 102} (2020)
  012005}.

\bibitem{gupta2021improving}
L.~Gupta, A.~Edelen, N.~Neveu, A.~Mishra, C.~Mayes and Y.-K.~Kim,
  \emph{Improving surrogate model accuracy for the lcls-ii injector frontend
  using convolutional neural networks and transfer learning}, {\emph{Machine
  Learning: Science and Technology} {\bfseries 2} (2021) 045025}.

\bibitem{kendall2017uncertainties}
A.~Kendall and Y.~Gal, \emph{What uncertainties do we need in bayesian deep
  learning for computer vision?}, {\emph{Advances in neural information
  processing systems} {\bfseries 30} (2017) }.

\bibitem{tagasovska2019single}
N.~Tagasovska and D.~Lopez-Paz, \emph{Single-model uncertainties for deep
  learning}, {\emph{Advances in neural information processing systems}
  {\bfseries 32} (2019) }.

\bibitem{antoran2020depth}
J.~Antor{\'a}n, J.~Allingham and J.M.~Hern{\'a}ndez-Lobato, \emph{Depth
  uncertainty in neural networks}, {\emph{Advances in neural information
  processing systems} {\bfseries 33} (2020) 10620}.

\bibitem{amodei2016deep}
D.~Amodei, S.~Ananthanarayanan, R.~Anubhai, J.~Bai, E.~Battenberg, C.~Case
  et~al., \emph{Deep speech 2: End-to-end speech recognition in english and
  mandarin},  in \emph{International conference on machine learning},
  pp.~173--182, PMLR, 2016.

\bibitem{shafighfard2024chained}
T.~Shafighfard, F.~Kazemi, F.~Bagherzadeh, M.~Mieloszyk and D.-Y.~Yoo,
  \emph{Chained machine learning model for predicting load capacity and
  ductility of steel fiber--reinforced concrete beams}, {\emph{Computer-Aided
  Civil and Infrastructure Engineering} (2024) }.

\bibitem{wahid2023multiphase}
M.F.~Wahid, R.~Tafreshi, Z.~Khan and A.~Retnanto, \emph{Multiphase flow rate
  prediction using chained multi-output regression models}, {\emph{Geoenergy
  Science and Engineering} {\bfseries 231} (2023) 212403}.

\bibitem{le2023comparing}
A.~Le~Borgne, X.~Marjou, F.~Parzysz and T.~Lemlouma, \emph{Comparing a
  composite model versus chained models to locate a nearest visual object},  in
  \emph{2023 IEEE/ACIS 8th International Conference on Big Data, Cloud
  Computing, and Data Science (BCD)}, pp.~231--236, IEEE, 2023.

\bibitem{drielsma2021scalableendtoenddeeplearningbaseddata}
F.~Drielsma, K.~Terao, L.~Dominé and D.H.~Koh, \emph{Scalable, end-to-end,
  deep-learning-based data reconstruction chain for particle imaging
  detectors},  2021.

\bibitem{buuck2023low}
M.~Buuck, A.~Mishra, E.~Charles, N.~Di~Lalla, O.~Hitchcock, M.~Monzani et~al.,
  \emph{Low-energy electron-track imaging for a liquid argon
  time-projection-chamber telescope concept using probabilistic deep learning},
  {\emph{The Astrophysical Journal} {\bfseries 942} (2023) 77}.

\bibitem{khek2022gamma}
B.~Khek, A.~Mishra, M.~Buuck and T.~Shutt, \emph{Gamma ray source localization
  for time projection chamber telescopes using convolutional neural networks},
  {\emph{AI} {\bfseries 3} (2022) 975}.

\bibitem{Abratenko_2022}
P.~Abratenko, R.~An, J.~Anthony, L.~Arellano, J.~Asaadi, A.~Ashkenazi et~al.,
  \emph{Novel approach for evaluating detector-related uncertainties in a
  lartpc using microboone data},
  \href{https://doi.org/10.1140/epjc/s10052-022-10270-8}{\emph{The European
  Physical Journal C} {\bfseries 82} (2022) }.

\bibitem{koh2023deep}
D.~Koh, A.~Mishra and K.~Terao, \emph{Deep neural network uncertainty
  quantification for lartpc reconstruction}, {\emph{Journal of Instrumentation}
  {\bfseries 18} (2023) P12013}.

\bibitem{koh2021evaluating}
K.D.H.~Koh, A.~Mishra and K.~Terao, \emph{Evaluating deep learning uncertainty
  quantification methods for neutrino physics applications},  in \emph{35th
  Conference on Neural Information Processing Systems}, 2021.

\bibitem{adams2020pilarnet}
C.~Adams, K.~Terao and T.~Wongjirad, \emph{Pilarnet: Public dataset for
  particle imaging liquid argon detectors in high energy physics}, {\emph{arXiv
  preprint arXiv:2006.01993} (2020) }.

\bibitem{wang2013fast}
S.~Wang and C.~Manning, \emph{Fast dropout training},  in \emph{international
  conference on machine learning}, pp.~118--126, PMLR, 2013.

\bibitem{petersen2024uncertainty}
F.~Petersen, A.~Mishra, H.~Kuehne, C.~Borgelt, O.~Deussen and M.~Yurochkin,
  \emph{Uncertainty quantification via stable distribution propagation},
  {\emph{International Conference on Learning Representations (ICLR)} (2024) }.

\bibitem{postels2019sampling}
J.~Postels, F.~Ferroni, H.~Coskun, N.~Navab and F.~Tombari, \emph{Sampling-free
  epistemic uncertainty estimation using approximated variance propagation},
  in \emph{Proceedings of the IEEE/CVF international conference on computer
  vision}, pp.~2931--2940, 2019.

\bibitem{shao2024uncertainty}
W.~Shao, J.~Xu, Z.~Cao, H.~Wang and J.~Li, \emph{Uncertainty-aware prediction
  and application in planning for autonomous driving: Definitions, methods, and
  comparison}, {\emph{arXiv preprint arXiv:2403.02297} (2024) }.

\bibitem{ivanovicimportance}
A.~Ivanovic, M.~Itkina, R.~McAllister, I.~Gilitschenski and F.~Shkurti,
  \emph{On the importance of uncertainty calibration in perception-based motion
  planning}, {\emph{IEEE International Conference on Robotics and Automation
  (ICRA)} (2024) }.

\bibitem{Calibration}
T.~Gneiting, F.~Balabdaoui and A.E.~Raftery, \emph{Probabilistic forecasts,
  calibration and sharpness},
  \href{https://doi.org/https://doi.org/10.1111/j.1467-9868.2007.00587.x}{\emph{Journal
  of the Royal Statistical Society: Series B (Statistical Methodology)}
  {\bfseries 69} (2007) 243}
  [\href{https://arxiv.org/abs/https://rss.onlinelibrary.wiley.com/doi/pdf/10.1111/j.1467-9868.2007.00587.x}{{\ttfamily
  https://rss.onlinelibrary.wiley.com/doi/pdf/10.1111/j.1467-9868.2007.00587.x}}].

\bibitem{mackay2003information}
D.J.~MacKay, \emph{Information theory, inference and learning algorithms},
  Cambridge university press (2003).

\bibitem{PhysRevD.104.072004}
{\scshape DeepLearnPhysics Collaboration} collaboration, \emph{Clustering of
  electromagnetic showers and particle interactions with graph neural networks
  in liquid argon time projection chambers},
  \href{https://doi.org/10.1103/PhysRevD.104.072004}{\emph{Phys. Rev. D}
  {\bfseries 104} (2021) 072004}.

\end{thebibliography}\endgroup

\end{document}